%% file: main.tex
\documentclass[manuscript]{acmart}
\newcommand{\commentout}[1]{}
\usepackage{enumerate}
\usepackage{enumitem}
\usepackage{multirow}
\usepackage{booktabs}
\usepackage{graphicx}
\usepackage{caption}
\usepackage{color,soul}
\usepackage{subcaption}
\usepackage{fontawesome5}
\usepackage{tabularray}
\usepackage{hyperref}
\usepackage[toc,page]{appendix}
\usepackage{colortbl}
\usepackage{tikz}
\usepackage{tabularx}
\usepackage{makecell}

\AtBeginDocument{%
  \providecommand\BibTeX{{%
    \normalfont B\kern-0.5em{\scshape i\kern-0.25em b}\kern-0.8em\TeX}}}

\begin{document}

\title[]{\textsc{Alloy}: Generating Reusable Agent Workflows from User Demonstration}

\author{Jiawen Li}
\affiliation{%
  \institution{University of Michigan, Ann Arbor}
  \city{Ann Arbor}
  \state{Michigan}
  \country{USA}
}

\author{Zheng Ning}
\affiliation{%
  \institution{University of Notre Dame}
  \city{Notre Dame}
  \state{Indiana}
  \country{USA}
}

\author{Yuan Tian}
\affiliation{%
  \institution{Purdue University}
  \city{West Lafayette}
  \state{Indiana}
  \country{USA}
}

\author{Toby Jia-jun Li$^{\star}$}
\affiliation{%
  \institution{University of Notre Dame}
  \city{Notre Dame}
  \state{Indiana}
  \country{USA}
}
\thanks{
\indent ~$^{\star}$ Co-corresponding.
}

\begin{abstract} 

Large language models (LLMs) enable end-users to delegate complex tasks to autonomous agents through natural language. However, prompt-based interaction faces critical limitations: Users often struggle to specify procedural requirements for tasks, especially those that don't have a factually correct solution but instead rely on personal preferences, such as posting social media content or planning a trip. Additionally, a ``successful'' prompt for one task may not be reusable or generalizable across similar tasks. We present Alloy, a system inspired by classical HCI theories on Programming by Demonstration (PBD), but extended to enhance adaptability in creating LLM-based web agents. \textsc{Alloy} enables users to express procedural preferences through natural demonstrations rather than prompts, while making these procedures transparent and editable through visualized workflows that can be generalized across task variations. In a study with 12 participants, \textsc{Alloy}’s demonstration-based approach outperformed prompt-based agents and manual workflows in capturing user intent and procedural preferences in complex web tasks. Insights from the study also show how demonstration-based interaction complements the traditional prompt-based approach.

\end{abstract}


\maketitle

\input{Files/1-Introduction}

\input{Files/2-RelatedWork}

\input{Files/3-System}
\input{Files/4-UserStudy}

\input{Files/5-Discussion}
\input{Files/6-FutureWork}
\input{Files/7-Conclusion}
\bibliographystyle{ACM-Reference-Format}
\bibliography{main}

\clearpage
\onecolumn

\appendix
\input{Files/Appendix.tex}
\end{document}

%% file: Files/1-Introduction.tex
\begin{figure}[htbp]
  \includegraphics[width=1\textwidth]{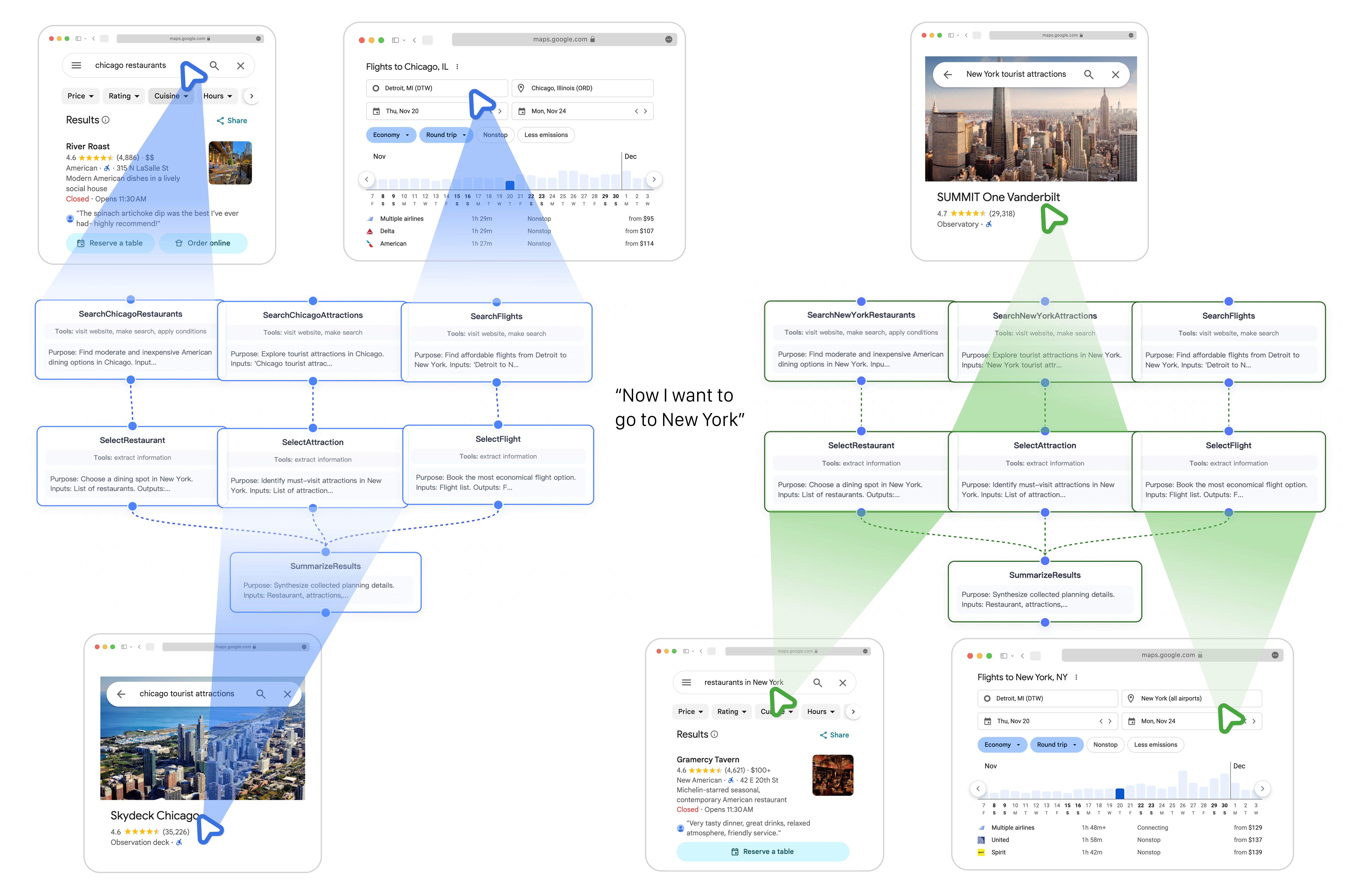}
  \caption{System overview of \textsc{Alloy}. User demonstrations (e.g. looking for tourist attractions, restaurants, and flight tickets) are collected as input for the multi-agent system, which dynamically generates task workflows that can be reused and generalized through natural language instruction.} 
  \label{fig:example}
\end{figure}

\section{Introduction}
For decades, human–computer interaction research has envisioned intelligent agents as collaborative partners capable of anticipating user needs, reduce cognitive burden, and simplify interaction with complex digital systems~\cite{maes1993agents,maes1994info,maes1997debate,zhang_agentic_2025,ning2025surveywebagentsnextgenerationai}. 
Recent advances in large language models (LLMs) have brought this vision closer to reality. 
LLM-based agents fundamentally outperform their rule-based predecessors~\cite{eliza1966, lieberman1995letizia} by leveraging extensive world knowledge, general instruction-following capabilities, and reasoning abilities for external tool use, enabling them to tackle more complex tasks with much greater flexibility and generalizability~\cite{Wang_2024}.


Among the various environments in which agents operate, the web stands out as one of the most challenging and universal ones. As most applications have transitioned to web-based platforms, users could perform a wide range of tasks on browsers, such as information seeking, document editing, collaborative design, and more.
Recent state-of-the-art web agents, such as Operator~\cite{openai2025operator} and BrowserUse~\cite{browser_use2024}, enable users to automate these tasks end-to-end using a simple text prompt that describes their goals. By leveraging the reasoning capabilities of underlying LLMs, these agents can autonomously plan and execute actions within a browser to help users complete tasks~\cite{zhou2024webarenarealisticwebenvironment, yang2025agenticwebweavingweb, ning2025surveywebagentsnextgenerationai}.



However, existing browser-use agents still face several key challenges.
First, users often struggle to clearly articulate their needs~\cite{qian-etal-2024-tell}, resulting in ambiguity and misalignment. 
In everyday use, people have already developed strong ``muscle memory'' for their routine web tasks. In other words, they know how to perform the task by themselves, while they may feel mentally demanding to describe what they do in detail in a natural language prompt~\cite{subramonyam2024promptCog, mahdavi2024}.
Second, most existing agent planning and execution processes are largely automated, making it challenging to maintain alignment with human intent. 
Without explicit mechanisms to explain and refine agent decisions, the execution results may easily contain mistakes that are difficult for humans to identify.
Third, many usage workflows are inherently similar and transferable. For example, paying an insurance fee can resemble paying an electricity bill. While user goals can evolve during interaction~\cite{pirolli2005sensemaking}, such workflows reflect underlying human values and preferences that are widely shared. However, existing systems require agents to plan and execute from scratch each time, ignoring opportunities to reuse and adapt prior knowledge, resulting in inefficiency and limited personalization.

To bridge the gap, this paper aims to address the question: \textit{How can users naturally externalize their procedural knowledge for web agents without requiring the explicit articulation of every task detail?}
We draw inspiration from classical HCI theories on \textit{Programming by Demonstration} (PBD), where users specify procedures through examples. 
Cognitive science research reveals that demonstration serves as a natural and efficient modality for conveying procedural knowledge. Humans readily demonstrate procedures when teaching~\cite{NIPS2016_b5488aef, Dautenhahn2002PBEasImitationGame}, because demonstration reduces cognitive burden on teachers by eliminating the need to verbalize tacit procedural knowledge~\cite{polanyi2009tacit}, while simultaneously providing learners with concrete, contextualized examples that ground abstract concepts in observable actions~\cite{clark1996}. For LLM-based agents, their strong understanding and reasoning abilities position them as capable learners who can infer goals, constraints, and intentions from observed demonstrations. What once challenged human learners in inferring the rationale behind demonstrated acts~\cite{sumers2020show} has become a strength for LLM-based agents, who can transform unspoken intention into grounded understanding.

Building upon these insights, we present \textsc{Alloy}\footnote{\textsc{Alloy} is an acronym for \textbf{A}gentic \textbf{L}ogic \textbf{L}earned from \textbf{O}bserving \textbf{Y}ou}, a system that transforms user demonstrations on browsers into editable and reusable LLM workflows that execute on a sequence of subtasks. Specifically, \textsc{Alloy} automatically updates the workflow while users are demonstrating their browser use. The workflow is visualized as a graph, where each node represents a ``subtask'' that is inferred automatically from the users' demonstrated actions. 
Behind each node, a dedicated browser agent is automatically created to execute the corresponding action sequence. 
Upon the visualized graph, users can directly edit the workflow by adding new nodes or removing existing nodes along with their dependent nodes, without considering action-level details (e.g., clicks or keystrokes). Additionally, users can customize the behavior of each sub-task agent corresponding to each node through natural language to better align the workflow with their preferences.
\textsc{Alloy} also allows users to save the generated workflows and reuse them as templates for future tasks. When faced with a new but related task, users can select a saved workflow, describe the new requirements in natural language, and \textsc{Alloy} will update the workflow to adapt to the new task while incorporating user preferences in early workflows.



We empirically evaluate \textsc{Alloy} through a within-subjects study (N=12) comparing demonstration-based, prompt-based, and manual workflows across three web tasks of varying complexity. The study focused on answering three key questions: (1) whether users can successfully generate workflows from demonstrations and generalize them to similar tasks with simple prompts; (2) how \textsc{Alloy}'s demonstration-based approach is when comparing to two baselines in capturing user intent and procedural preferences; and (3) what challenges users encountered and how they leveraged \textsc{Alloy}'s visual editing interface to address them. 
Participants completed each task under three conditions: using \textsc{Alloy}'s full demonstration-based system, manually constructing workflows without AI assistance (Manual Baseline), and using a single LLM-based agent controlled through natural language prompts (LLM Baseline). For the generalization evaluation, participants used their \textsc{Alloy}-generated workflows to complete structurally similar task variants with single prompts. Through quantitative usability assessments and semi-structured interviews, we gathered insights into user perceptions, preferences, and interaction patterns when using demonstrations versus prompts to specify procedural requirements for LLM-based agents. We find that users prefer demonstrations to natural language prompts or manually constructed workflows in exploratory tasks with implicit goals. Based on this result, we propose design implications for procedure-aligned human agent systems. 

In summary, this paper presents the following contributions:
\begin{itemize}
    \item \textsc{Alloy}, a browser-based agentic system that transforms user demonstrations into editable and reusable workflows, enabling lay users to generate, adapt, and generalize LLM-based agent workflows.
    \item A comprehensive evaluation of \textsc{Alloy}’s workflow generation and generalization capabilities, compared to a manual workflow editing baseline and a prompting-based agent baseline. The study also includes the analysis of user perceptions, preferences, and behaviors.
    \item Design implications for developing agentic systems that more effectively align with user intent and preferences through procedural demonstrations.
\end{itemize}

%% file: Files/2-RelatedWork.tex
\section{Related Work}


\subsection{Programming by Demonstration for Task Automation}




Programming by Demonstration (PBD) allows non-programmers to automate activities by performing actions directly, with systems generating reusable programs from these demonstrations~\cite{myers1986,lau_programming_1998}. PBD has been successfully applied across diverse domains, including UI construction~\cite{myers1986pbdui}, text editing~\cite{miller2002lapis}, mobile applications~\cite{toby2017sugilite}, and web automation~\cite{miwa,hartmann2007,PbdSarah,pbd2022,CoScripter,rebeccaScrapeViz,multi-click}.
The web environment presents unique challenges for PBD due to dynamic content, inconsistent interface structures, and cross-site workflows~\cite{whyPBDfails}. Early web PBD systems such as CoScripter~\cite{CoScripter} and Vegemite~\cite{Lin2009Vegemite} demonstrated the potential of demonstration-based web automation but struggled with brittleness when encountering interface changes. Recent systems have attempted to improve robustness and user control: Rousillon~\cite{rousillon} addresses brittleness by providing a Scratch-style visual syntax for editing demonstrations, while MIWA~\cite{miwa} improves understandability through natural language explanations and visual feedback. Despite these advances, these systems primarily rely on rule-based heuristics to map demonstrations to actions, limiting their ability to generalize across inconsistent web interfaces and content variations. Furthermore, while user demonstrations encode rich information about intents and procedural preferences, existing PBD systems focus primarily on capturing structural patterns rather than the semantic meaning behind user actions.

Recent work has explored capturing richer information from user demonstrations beyond raw operations. Yin et al.~\cite{yin2025taskmind} propose automatically modeling cognitive dependencies between GUI operations to improve task generalization. However, their approach represents workflows as linear action sequences with sequential dependencies and relies on replaying recorded GUI operations, which struggles with the dynamic and inconsistent nature of web interfaces. In contrast, \textsc{Alloy} addresses task-level procedural alignment by transforming demonstrations into graph-structured workflows where each node represents an LLM-based sub-task agent rather than atomic operations. This design provides two key advantages: (1) graph structures enable complex procedural patterns including conditional branches, parallel sub-tasks, and iterative processes that cannot be expressed in linear sequences, and (2) LLM-based agents at each node can adapt to interface variations and dynamic web content by reasoning about semantic goals rather than replaying fixed GUI operations. Together, these capabilities enable \textsc{Alloy} to handle complex, multi-step web workflows that require both procedural flexibility and robustness to interface changes.


\subsection{LLM-based Web Agents}
Early web agents relied on rule-based systems or reinforcement learning to automate web interactions. Systems like Letizia~\cite{lieberman_autonomous_1997} and CoScripter~\cite{CoScripter} used scripting languages to specify and execute web automation tasks, while learning-based agents such as WoB~\cite{tianlin2017wob} learned policies in simplified environments. However, these agents suffered from brittleness to interface changes, limited generalization beyond training scenarios, and required either extensive manual engineering or large task-specific datasets~\cite{Xiang2022DOM,gur2018learning}, making them inaccessible to end-users.

The emergence of large language models has fundamentally transformed the capabilities of web agents. Modern LLM-based agents such as Operator~\cite{openai2025operator}, BrowserUse~\cite{browser_use2024}, and WebVoyager~\cite{he2024webvoyagerbuildingendtoendweb} leverage pre-trained language understanding to interpret natural language instructions, reason about web content semantically, and generalize across diverse websites without task-specific training. Despite these autonomous capabilities, current LLM-based agents face critical challenges in procedural alignment and reusability. They optimize for task completion rather than user-preferred execution paths~\cite{zhang2025characterizingunintendedconsequenceshumangui, tang2025darkpatternsmeetgui, Mininger_Laird_2018}, causing misalignment when procedural preferences matter (e.g., preferring direct flights over cheapest options). Recent attempts through prompt engineering~\cite{ye-etal-2024-prompt} and few-shot learning~\cite{verma-etal-2025-adaptagent}, require users to articulate tacit procedural knowledge explicitly, which is inherently difficult as such knowledge is often implicit and action-based rather than declarative~\cite{polanyi2009tacit}. Furthermore, successful executions remain confined to single sessions with little reusability; even when traces are exposed, they consist of low-level actions rather than transferable high-level strategies.

\textsc{Alloy} addresses these limitations through demonstration-driven workflow generation. Rather than relying on abstract prompts, \textsc{Alloy} enables users to demonstrate their procedures directly, transforming demonstrations into editable, graph-structured workflows where each node represents an LLM-powered sub-task agent. This design achieves procedural alignment by grounding workflows in actual user demonstrations while maintaining LLM adaptability, and enables reusability by representing workflows as visual, editable templates that users can generalize to similar tasks through simple natural language prompts.

\subsection{Workflow Representation for Human Understanding and Agent Execution}

Representing workflows has long been central to both human cognition and agent system design, serving as a bridge between how people conceptualize procedures and how computational systems execute them.

\subsubsection{Human Workflow Representation and Understanding}
Workflows and task models externalize procedural knowledge, enabling people to decompose complex activities into hierarchical steps~\cite{STANTON20065Hierarchicaltask}, reason about dependencies~\cite{sweller1988cognitive}, and communicate execution strategies to others~\cite{SCAIFE1996185howdographicalrepresentationswork}. This perspective is further supported by cognitive script theory~\cite{schank2013scripts} (which models stereotyped action sequences) and by HCI work on task models (e.g., ConcurTaskTrees~\cite{taskTree1997}), as well as theories of distributed cognition that emphasize the role of external representations in collaborative and individual problem solving~\cite{norman2013design}. Later work in end-user programming and workflow authoring tools extended this idea, allowing users to specify workflows through direct manipulation and example-based interactions~\cite{hartmann2007chi, tim20205ply, zhang_chainbuddy_2025}. Such systems highlight that users prefer concrete, situated representations that align with their own procedural reasoning rather than abstract programmatic logic~\cite{GREEN1996131}. These findings motivate our design of a workflow representation of user demonstrations, making the procedural logic both interpretable to humans and operational for AI agents.

\subsubsection{Agent Workflow Construction for Complex Tasks}
As LLM-based agents have grown more capable, recent work has increasingly adopted multi-agent workflows that decompose complex tasks into specialized sub-agents. This architectural shift offers several advantages: improved reliability in solving complex problems~\cite{hu_automated_2025}, modularity that facilitates error handling~\cite{niu2025flow}, and more predictable resource usage~\cite{tan_span_2025}.

Current approaches to building agent workflows fall into three main paradigms. Developer-centric frameworks such as LangChain~\footnote{https://github.com/langchain-ai/langchain} and AutoGen~\cite{wu2023autogenenablingnextgenllm} provide powerful orchestration capabilities. Low-code visual platforms like Dify~\cite{leatherwood2025dify}, Coze~\cite{coze}, and n8n~\cite{n8n} lower technical barriers with visual, block-based interfaces. More recent systems like AFlow~\cite{zhang_aflow_2025} and SPAN~\cite{tan_span_2025} automatically generate or optimize workflows from natural language task descriptions, reducing manual specification burden.

However, these approaches share critical limitations. Code-based  require programming expertise to manually code workflow logic. Low-code platforms still leave the challenge of end-user programming~\cite{learningbarriers2004} unsolved by assuming users can explicitly translate their procedural knowledge into logical steps. Automated generation methods struggle to capture user-specific execution strategies from abstract textual descriptions. Furthermore, automatic workflow generation has been largely confined to closed-domain, static tasks such as Q\&A or mathematical problem-solving, and has yet to address open-ended, interactive environments like web-based tasks. \textsc{Alloy} addresses these gaps by enabling end-users to specify agent workflows through direct demonstration in situ, naturally preserving their preferred execution strategies while maintaining the adaptability of LLM-based agents.


%% file: Files/3-System.tex
\section{\textsc{\textsc{Alloy}} System}
The \textsc{Alloy} system transforms user demonstrations into editable and reusable LLM workflows through an integrated browser-based environment. Section~\ref{sec:designgoals} outlines the design principles that guide \textsc{Alloy}. Section~\ref{sec:scenario} provides a usage example that illustrates how users demonstrate, edit, and generalize workflows, utilizing the detailed features discussed in Section~\ref{sec:features}. Section~\ref{sec:pipeline} explains the multi-agent technical pipeline supporting workflow generation, adaptation, and execution with implementation details in Section~\ref{sec:implementation}.

\subsection{Design Goals}\label{sec:designgoals}
We conducted a comprehensive literature review and analysis of the limitations of existing agentic systems~\cite{cao_spider2-v_2024,horvitz_principles_1999,terveen_helping_1996,maes1997debate,mahdavi2024} and identified three key design goals for \textsc{Alloy}:

\textbf{DG1: Enable procedural specification through demonstration.}
Users often struggle to articulate task procedures through natural language, especially when tasks involve many steps and complex conditional logic. 
To enhance the efficiency of alignment between humans and agents, \textsc{Alloy} should allow users to intuitively demonstrate their preferred execution strategies through direct actions in the browser, avoiding the cognitive burden of translating procedural knowledge into textual descriptions.

\textbf{DG2: Provide transparent, editable workflow representations at the task level.}

AI agents inevitably make errors during workflow generation, requiring users to inspect and correct their outputs~\cite{zhang2025characterizingunintendedconsequenceshumangui, cai-etal-2024-low-code}. However, action-level representations (e.g., low-level API calls) are difficult to interpret and modify without technical expertise. To address this, \textsc{Alloy} should visualize workflows to align with users' mental models. In the meantime, users should be able to easily manipulate the workflow without programming expertise, supporting understanding, agency, and effective error-handling.


\textbf{DG3: Support workflow reuse across relevant tasks.}
Users frequently encounter structurally similar tasks (e.g., booking different trips, researching different products) where effective strategies could be reused. 
\textsc{Alloy} should enable users to save demonstrated workflows as templates and easily adapt them to task variations (e.g., natural language instructions), thereby avoiding redundant effort in re-demonstrating similar procedures.

\subsection{Example Usage Scenario}\label{sec:scenario}

\begin{figure}
    \centering
    \includegraphics[width=1\linewidth]{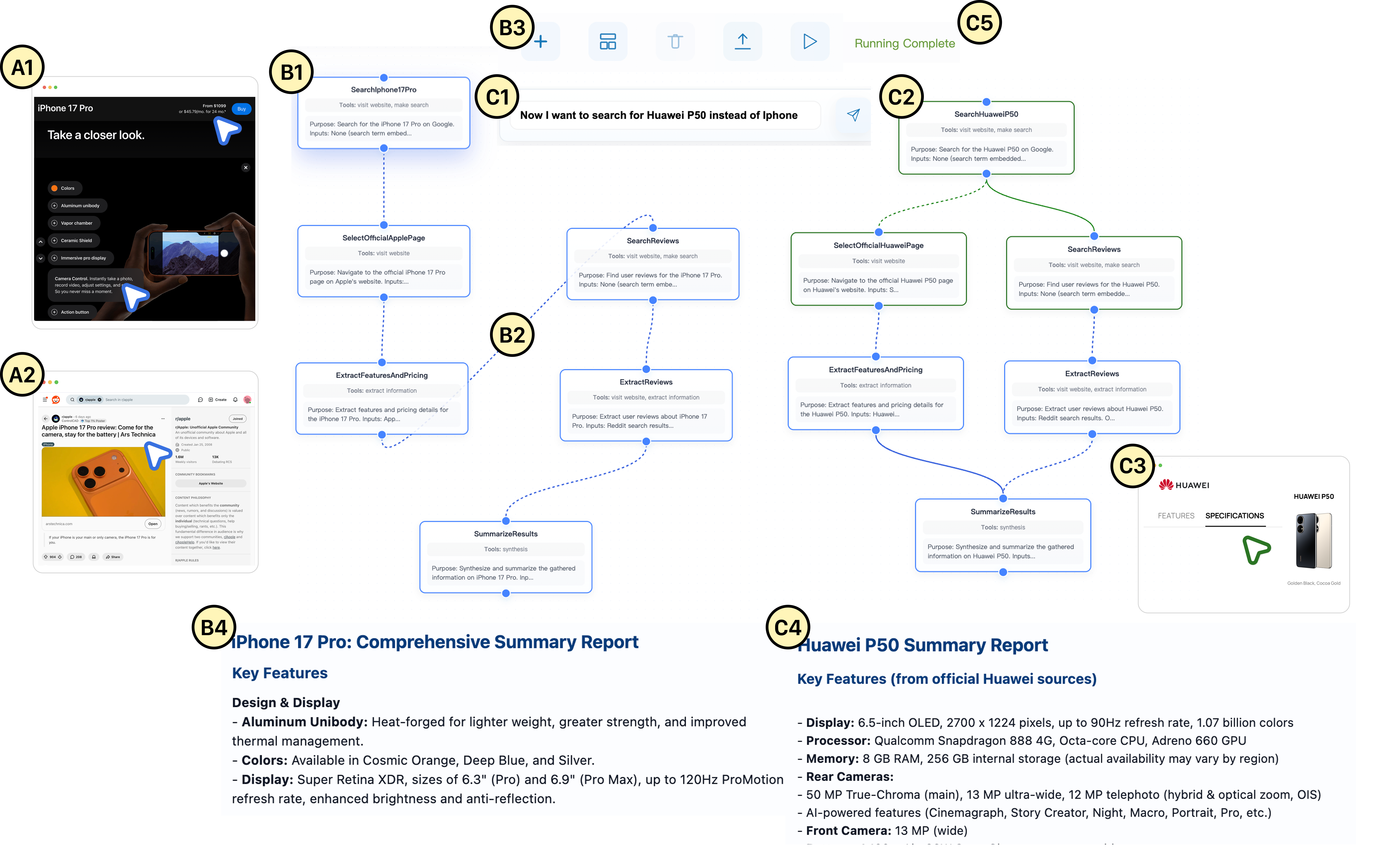}
    \caption{UI overview of \textsc{Alloy}. Users demonstrate a task by performing it directly in the browser (A1-A2), from which \textsc{Alloy} automatically generates a structured workflow (B1). Users can refine the workflow through visual editing (B2), execute it via the control panel (B3), and review execution outcomes (B4). To adapt the workflow to new scenarios, users provide natural language instructions (C1), then execute the generalized workflow (C2) with real-time monitoring (C3) to obtain new results (C4). The status panel (C5) provides continuous feedback on system state throughout the interaction.}
    \label{fig:UI}
\end{figure}

This section presents an example usage scenario in which Alice uses \textsc{Alloy} to create a reusable workflow to compare cell phone specifications, prices, and reviews across multiple brands. 


\textbf{Demonstrating the Research Process.} Alice is shopping for a new cell phone and needs to compare information across multiple brands, which can be a tedious process to repeat manually. Using \textsc{Alloy}, she opens her browser with the \textsc{Alloy} extension activated. A panel displaying the task-level workflow will open automatically and update in real time as Alice demonstrates each step. 

She simply follows her regular way of researching different models: she begins by searching ``iPhone 17 Pro'' on Google, clicking the link to Apple's official website to examine design specifications and pricing, then performing another search for ``iPhone 17 Pro reviews Reddit'' to read user experiences. As Alice interacts with each webpage, \textsc{Alloy}'s demonstration recording system (Section~\ref{f1}) captures her browser actions along with contextual DOM information, while the real-time workflow generation system (Section~\ref{f2}) continuously analyzes these interactions in the background, inferring her research goals and procedural structure.

\textbf{Reviewing and Refining the Generated Workflow.} After completing her demonstration, Alice shifts her attention to the workflow visualization panel, where six nodes have appeared representing the inferred sub-tasks as shown in Figure~\ref{fig:UI}. Each node displays a concise natural language description of its purpose, making the workflow structure immediately comprehensible at the task level rather than showing action macros. Alice notices that nodes for gathering specifications and reading reviews are connected sequentially, but she believes these tasks should run in parallel since they're independent. Using the visual workflow editing interface (Section~\ref{f3}), she rearranges the connection logic. She then hovers over node 2 to verify its description accurately captures her intent, and finally clicks the ``Export Workflow'' button to store this template for future use.

\textbf{Validating Through Execution.} To verify the workflow behaves as intended, Alice clicks the``Execute Workflow'' button. The browser-level workflow execution system orchestrates the automated browser actions with the specification gathering and review collection running simultaneously as she designed. Within seconds, a results panel slides in from the right, displaying structured output with the iPhone 17 Pro's specifications, pricing, and synthesized Reddit insights organized in an easy-to-scan format. Satisfied that the workflow accurately captures her research process, Alice confirms it's ready for reuse.

\textbf{Generalizing to New Products.} Alice now leverages \textsc{Alloy}'s generalization capability for other phone brands. She clicks on the saved workflow card and types in the generalization text box at the top of the panel: ``now I want to search for Huawei Mate 60 Pro instead of iPhone'' \textsc{Alloy}'s workflow adaptation system (Section~\ref{f4}) processes this prompt, identifying product-specific parameters (brand name and model) in the workflow and replacing them with the new values while preserving the overall research structure. The node descriptions update in real-time to reflect ``Huawei Mate 50'' instead of ``iPhone 17 Pro.'' Alice clicks ``Execute Workflow'' again and, within seconds, receives comprehensive Huawei information. She continues adapting the workflow by typing new prompts—``Samsung Galaxy S24 Ultra,'' ``Google Pixel 9 Pro''—and executing each adaptation. Within minutes, she collects consistently-formatted comparison data across multiple brands, transforming what would have been repetitive manual research into a streamlined, reusable workflow created through natural demonstration and refined through intuitive visual editing.

\subsection{System Features}\label{sec:features}
 
\textsc{Alloy}'s design incorporates four key features that address the challenges of procedural alignment and workflow reuse in LLM-based web automation.

\subsubsection{F1: Demonstration-Based Workflow Generation}\label{f1}. Rather than requiring users to explicitly describe task procedures through natural language, \textsc{Alloy} captures user demonstrations as they naturally perform web tasks. This approach reduces cognitive burden by eliminating the need to translate tacit procedural knowledge into textual descriptions—users simply perform tasks as they normally would, and \textsc{Alloy} infers the underlying workflow structure from observed actions. The system selectively logs meaningful browser interactions (clicks, text input, form submissions, navigation) along with their contextual metadata (DOM attributes, element labels, page content) to construct a rich representation of user intent. This demonstration-based approach addresses a fundamental limitation of prompt-based agents: the difficulty of articulating complex, multi-step procedures that involve implicit decision-making and contextual dependencies.

\subsubsection{F2: Task-Level Visual Workflow Representation.} \label{f2}
\textsc{Alloy} visualizes generated workflows as node graphs displayed in the browser's side panel, where each node represents a semantically meaningful sub-task rather than low-level browser operations. This task-level abstraction serves two critical purposes: (1) it reduces cognitive load by allowing users to understand workflow logic without inspecting every granular action, and (2) it provides a natural unit of manipulation for workflow editing. Each node encapsulates a configurable LLM-powered agent with its own description, prompt, and available tools, while edges represent data flow and logical dependencies between sub-tasks. This visual representation makes the workflow's procedural structure transparent and comprehensible, addressing users' need to understand and trust agent behavior before deployment.

\subsubsection{F3: Direct Manipulation Editing Interface}\label{f3}. Users maintain full control over generated workflows through direct manipulation: they can add new nodes to extend functionality, delete nodes to remove unnecessary steps, reconnect edges to modify task dependencies, or re-record individual nodes to demonstrate alternative procedures. Each node's configuration—including its natural language prompt and tool access—can be modified directly to refine agent behavior without regenerating the entire workflow. This editing capability addresses two key challenges: (1) it enables users to correct misalignments between generated workflows and their intended procedures, and (2) it supports iterative refinement as users discover edge cases or evolving requirements. By providing editing capabilities at the task level rather than the code level, \textsc{Alloy} makes workflow customization accessible to non-programmers while preserving the adaptability of LLM-based agents.

\subsubsection{F4: Prompt-Based Workflow Generalization}\label{f4}. Once users create and validate a workflow, \textsc{Alloy} enables reuse across similar tasks through simple natural language instructions. Rather than re-demonstrating similar procedures, users can select a saved workflow and describe how the new task differs (e.g., "search for laptops instead of phones" or "plan a trip to Paris instead of Tokyo"). \textsc{Alloy}'s two-agent adaptation pipeline automatically identifies task-specific parameters in the original workflow, replaces them with semantic placeholders, and instantiates these placeholders according to the user's new requirements. This generalization mechanism transforms demonstrated workflows from single-use scripts into reusable templates, significantly reducing the effort required for structurally similar tasks while maintaining alignment with users' preferred execution strategies.

\subsection{Technical Pipeline}\label{sec:pipeline}

\begin{figure}
    \centering
    \includegraphics[width=1\linewidth]{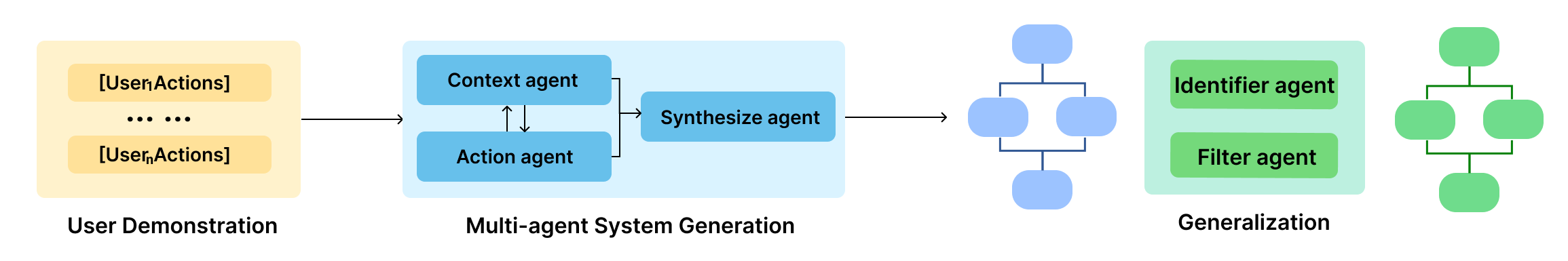}
    \caption{\textsc{Alloy}'s technical pipeline. \textsc{Alloy} collects user demonstration as inputs for a multi-agent system to generate the corresponding workflow. The workflow can be further adapted according to natural language prompt via a two-agent pipeline.}
    \label{fig:UI}
\end{figure}

\textsc{\textsc{Alloy}} is a multi-agent system with a visual workflow interface. In this section, we introduce the underlying technical pipeline of \textsc{\textsc{Alloy}}, which supports demonstration recording, direct manipulation editing, workflow generation, generalization, and execution. 


\subsubsection{User Demonstration Collection}
To infer users’ implicit task goals and intentions, we collect a set of common browser \textit{actions}, including clicks, text selection, text input, form submissions, and URL navigations. Alongside each action, we record the \textit{contextual metadata}, such as the class, tag, and value attributes of the corresponding DOM elements, to capture the semantic implications of the interaction. This approach is also generalizable because it relies on standard DOM APIs (\texttt{elementFromPoint}, \texttt{event.target}, and \texttt{FormData}) that are available in all mainstream browsers. 

To avoid user noise from meaningless or repeated actions, we selectively log meaningful user actions by filtering out low-information elements (e.g., background containers, generic layout divs, raw input placeholders), postponing event capture to debounce transient input changes, and recording only finalized interactions with minimal contextual metadata.


\subsubsection{Task-Level Workflow Generation and Adaptation}
We designed a three-agent pipeline that transforms user interaction logs into structured JSON workflows. The context agent extracts user goals, interests, and constraints from demonstration metadata while preserving all concrete values exactly as demonstrated. In parallel, the action agent processes the interaction log to identify high-level action sequences and task phases, applying heuristics to reconstruct completed inputs and group related events. The synthesizer agent then combines these analyses to generate a task-level workflow graph where each node represents a semantically meaningful sub-task rather than low-level operations. Each node includes a name, dependency structure (parent-child lists), required tools, and a detailed natural language prompt that specifies the purpose, inputs, outputs, preconditions, execution steps, and contingency handling. As the user continues interacting with the system, the workflow dynamically evolves to reflect new actions and intentions. The prompts for the agents can be found in Appendix~\ref{app:generation}.

The system achieves workflow adaptation through a sophisticated two-agent pipeline that enables generalization from specific executions to reusable templates. When users provide natural language instructions for workflow modification, the first Identifier agent performs semantic abstraction by analyzing the executed workflow and systematically replacing task-specific literals, such as URLs, product names, CSS selectors, and concrete values, with semantic placeholders. This agent maintains the workflow's structural integrity while documenting each placeholder with its semantic description. The second Filter agent then instantiates these placeholders according to the user's intent, leveraging both explicit instructions and contextual inference to generate concrete values aligned with the new task requirements. The prompts for the agents can be found in Appendix~\ref{app:generalization}.

\subsubsection{Browser-Level Workflow Execution}

The workflow execution process begins when the user exports their satisfied workflow as a JSON file, which is then automatically converted into an executable Python script using a predefined code template. Each node in the workflow is transformed into an asynchronous function that instantiates an LLM agent integrated with the Playwright MCP server for browser automation. All node functions share a common SQLiteSession that persists their execution results, enabling each agent to access the complete interaction history from previous nodes and maintain continuity throughout the workflow.

The workflow itself is structured as a \textit{Directed Acyclic Graph} and executed in topological order using Kahn's algorithm. The execution engine first identifies root nodes—those with no parent dependencies—and initializes them in an execution queue. To maximize efficiency, all nodes at the same dependency level are executed concurrently. Nodes at subsequent levels only begin execution once all their parent nodes have completed, ensuring proper dependency resolution while maintaining maximum parallelism.

During execution, the system maintains a shared results dictionary that stores the output of each completed node. The execution continues level by level until all nodes have been processed. Finally, the output from the last executed node is extracted and sent to the backend as the final workflow result, providing users with the complete outcome of their multi-step agent workflow.

\subsection{Implementation}\label{sec:implementation}
\textsc{Alloy} is implemented as a Chrome browser extension with a three-tier architecture. The frontend, built with React\footnote{https://react.dev/} and ReactFlow\footnote{https://reactflow.dev/}, provides interactive workflow visualization and uses the Chrome Extension API to capture user demonstrations. The Flask-based backend orchestrates workflow generation and adaptation using the LangGraph\footnote{https://www.langchain.com/langgraph} framework with GPT-4o-configured agents for intent inference and task abstraction. During execution, each workflow node instantiates an independent agent configured with GPT-4.1, which interfaces with Chrome browsers through the Playwright MCP\footnote{https://github.com/microsoft/playwright-mcp} server via the Chrome DevTools Protocol (CDP) to perform automated web interactions.

%% file: Files/4-UserStudy.tex
\section{User Study}
We conducted a lab user study to evaluate the usability, effectiveness, and usefulness of \textsc{\textsc{Alloy}} in real-world web tasks and to understand how people externalize, refine, and generalize their procedural knowledge through demonstration-based interaction with the system.
The study aims to answer the following questions:




\begin{itemize}
    \item \textbf{RQ1:} How does \textsc{\textsc{Alloy}}'s demonstration-based approach compare to baseline methods in terms of task completion effectiveness and usability?
    
    \item \textbf{RQ2:} What are users' experiences and perceptions of \textsc{\textsc{Alloy}}'s design features? Specifically, demonstration-based workflow generation, visual editing capabilities, and error handling.
    
    \item \textbf{RQ3:} How do users' mental models evolve using \textsc{\textsc{Alloy}}? Specifically, how do demonstrations complement prompts in expressing users' intent?
\end{itemize}

\subsection{Participants}
We recruited 12 participants (denoted P1-P12) through social media and word of mouth. Our participants (6 men, 6 women) were aged 20-24. 11 of 12 (91.7\%) participants had prior experience of using web agents, and most participants (75\%) reported frustration that LLMs sometimes act beyond their intended instructions, while 16.7\% found it difficult to precisely describe desired actions. A smaller portion (8.3\%) expressed interest in building their own LLM agent systems but lacked the knowledge to do so. We compensated each participant with a \$25 USD Amazon gift card for their time.

\subsection{Study Design}
Each study session lasted about 60 minutes. The studies were conducted in person. All sessions were screen- and voice-recorded. Participants were provided with a consent form, which they reviewed and signed before the study. The study employed a within-subjects design, in which each participant experienced three conditions for three web tasks. The study procedure is detailed in Section 4.2.3. Our study protocol was approved by our Institutional Review Board (IRB). 

\subsubsection{Tasks}
We designed three representative daily web tasks (see Table~\ref{tab:tasks}) spanning varying levels of complexity: social media content posting (low difficulty), trip planning (medium difficulty), and news information aggregation (high difficulty). Task difficulty was determined based on three criteria: (1) number of required action steps, (2) number of information sources or websites to be accessed, and (3) overall workflow complexity. To minimize priming effects, tasks were presented as naturalistic scenario descriptions rather than explicit procedural instructions, and participants were instructed not to copy task descriptions verbatim when creating prompts. To evaluate \textsc{\textsc{Alloy}}'s generalizability, each task was paired with a structurally similar variant that maintained the same procedural structure while varying surface-level content (e.g., different destinations for trip planning). Complete task descriptions are provided in the Appendix~\ref{app:task}. Participants completed all three tasks, each under a different condition, with task order counterbalanced across participants to control for learning and fatigue effects.

\begin{table}[t]
\centering
\begin{tabular}{p{3cm}p{6.5cm}p{4cm}c}
\toprule
\textbf{Task Type} & \textbf{Scenario Description} & \textbf{Typical Sources / Websites} & \textbf{Difficulty} \\
\midrule
\textbf{Social Media Content Creation} & 
Compose and cross-post personalized travel reflections across multiple platforms, adapting tone and format for different audiences. &
X (Twitter), Threads. &
Low \\
\midrule
\textbf{Trip Planning} & 
Plan a weekend itinerary (e.g., Chicago or New York) including attractions, restaurants, and transport, while comparing flights and mapping routes. &
Google Maps, TripAdvisor, airline booking sites. &
Medium \\
\midrule
\textbf{Information Aggregation} & 
Investigate recent H1B or STEM OPT policy changes by synthesizing credible updates from official, professional, and community sources. &
Google search, USCIS, news media, LinkedIn, immigration forums, Reddit. &
High \\
\bottomrule
\end{tabular}
\caption{Overview of user study tasks with varying complexity levels. Difficulty was determined by the number of required steps, sources to visit, and workflow complexity.}
\label{tab:tasks}
\end{table}

\subsubsection{Conditions}
The study employed a within-subjects design with three conditions, all implemented within the \textsc{\textsc{Alloy}} interface:

\begin{itemize}
\item Handcrafted Workflow (Manual Baseline): Participants manually constructed workflows using \textsc{\textsc{Alloy}}'s visual node editor without demonstration or agent assistance. This condition isolated the effects of manual workflow specification.

\item Single-Agent Prompting (LLM Baseline): Participants used a single workflow node containing an LLM-based web agent that was controlled entirely through natural language prompts. This condition replicated standard prompt-based agent interaction, as each \textsc{\textsc{Alloy}} node functions as an independent agent.

\item \textsc{\textsc{Alloy}} (Full System): Participants used the complete \textsc{\textsc{Alloy}} system, including demonstration-based workflow generation, visual editing capabilities, and prompt-based generalization for task variants.
\end{itemize}

\subsubsection{Study Procedure}
Each study session began with informed consent and the collection of demographic data. The experimenter then provided a high-level system overview (5 minutes), followed by a hands-on tutorial (10 minutes) using an example task of searching for information about the latest iPhone. During this tutorial, participants practiced the complete \textsc{Alloy} workflow by demonstrating actions to generate a workflow, editing the generated workflow, executing the workflow, and generalizing it to a variant task using a single natural language prompt. Participants were encouraged to interact with, export, and execute workflows themselves to build familiarity with the interface. Following the tutorial, participants completed all three web tasks under different task conditions, with each task taking approximately 10--15 minutes, including task explanation and participants operation. For each task, participants were asked if the outcome of the workflow was satisfactory. If not, users can continue their attempts until they are satisfied. Users are asked to complete the task pair only for tasks with full \textsc{Alloy} condition to assess the adaptability of \textsc{Alloy}'s system.
After completing the tasks, participants evaluated each condition using a 7-point Likert scale questionnaire assessing usability, usefulness, and interaction design preferences. Finally, a semi-structured interview explored participants' overall experiences, gathered detailed feedback on \textsc{Alloy}'s features and output quality, and elicited suggestions for improvement and perceived benefits.

\section{Results}
We present our findings in three parts according to our research questions. First, we report quantitative measures of task completion effectiveness and usability, comparing \textsc{\textsc{Alloy}}'s demonstration-based approach against baseline methods (RQ1). Second, we examine users' experiences with \textsc{\textsc{Alloy}}'s key design features, including the demonstration workflow, visual editing interface, and error handling mechanisms through qualitative analysis of user feedback and observed interaction patterns (RQ2). Third, we trace the evolution of users' mental models as they progressed through the demonstration, workflow generation, and generalization phases (RQ3).

\subsection{System Effectiveness and Usability}
\subsubsection{Task Completion and Attempts}
We analyzed first-attempt completion rates across the three conditions (Table~\ref{table:task-complete}). Overall, participants achieved satisfactory outcomes on their initial attempt in 75\% (9/12) of tasks using \textsc{Alloy}, 83\% (10/12) using manually crafted workflows, and 58\% (7/12) using the single-agent LLM baseline. For initially unsuccessful tasks, participants typically achieved satisfactory results within three additional attempts. One participant in the LLM baseline condition abandoned the task without completion. For generalization with full \textsc{Alloy}, all 12 successfully used natural language prompt to adapt the workflow to the task alternative within two attempts. 

\begin{table}[h!]
\centering
\begin{tabular}{lccc}
\hline
Task & \textsc{\textsc{Alloy}} & LLM Baseline & Manual Workflow \\
\hline
Task 1 & 4/4 & 1/4 & 4/4 \\
Task 2 & 2/4 & 3/4 & 3/4 \\
Task 3 & 3/4 & 3/4 & 3/4 \\
\textbf{Overall }& \textbf{9/12} & \textbf{7/12} & \textbf{10/12} \\
\hline
\end{tabular}
\caption{Comparison of task completeness across \textsc{Alloy}, LLM Baseline, and Manual Workflow}
\label{table:task-complete}
\end{table}


\subsubsection{Subjective User feedback}\label{result:survey}
\begin{figure}[htbp]
  \includegraphics[width=1\textwidth]{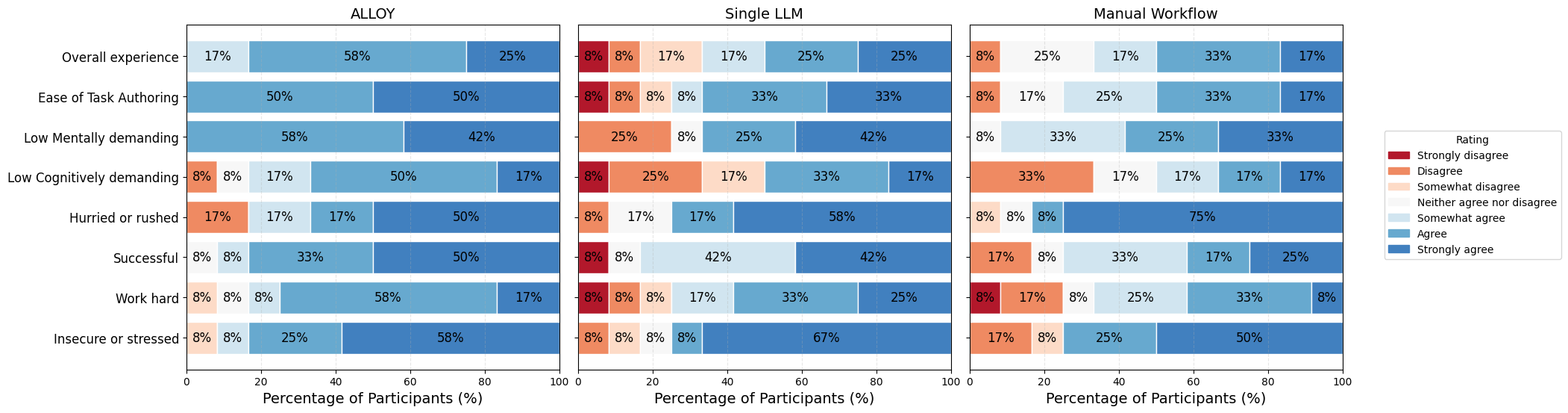}
  \caption{Overall comparison of user ratings of NASA-TLX survey for \textsc{Alloy} under three conditions.} 
  \label{fig:nasatlx}
\end{figure}

The overall subjective user feedback across three different difficulty-level tasks with three conditions is shown in Fig~\ref{fig:nasatlx}, \textsc{Alloy} received higher ratings for overall performance and self-reported success compared to manual workflow and single LLM-based agents. \textsc{Alloy} also has lower mental and cognitive demands.

\begin{figure}[htbp]
  \includegraphics[width=1\textwidth]{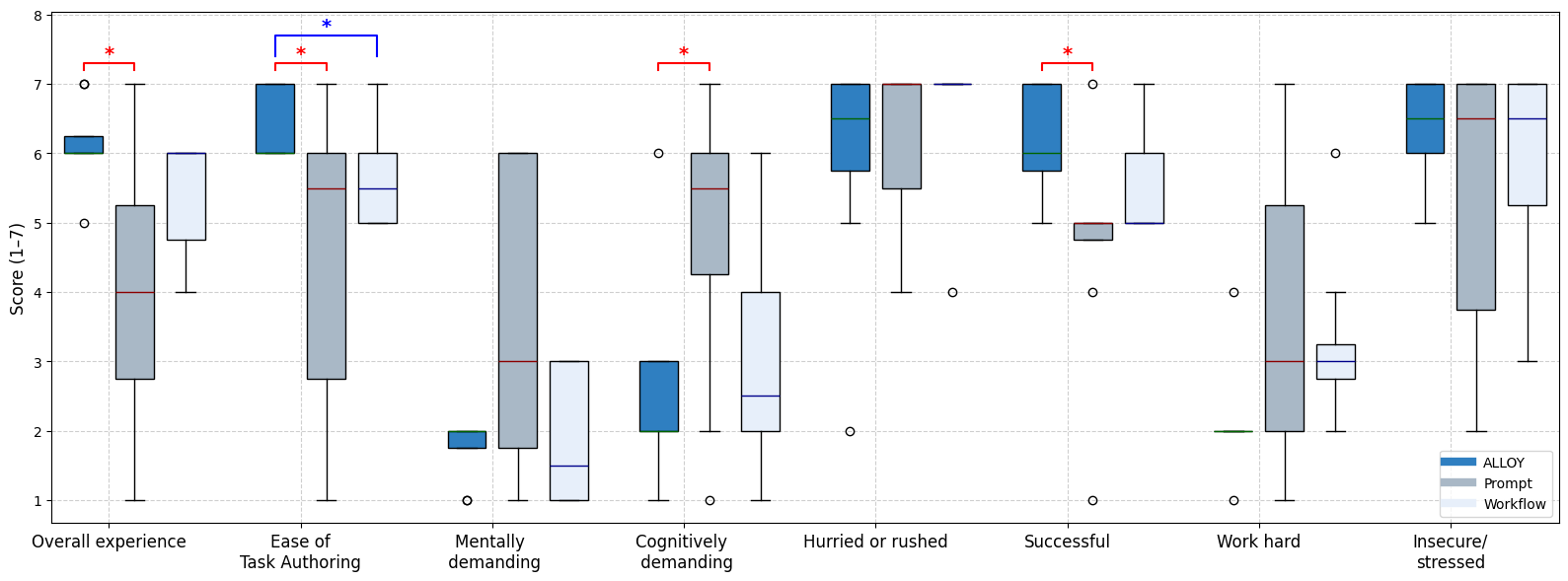}
  \caption{Comparison of user ratings of NASA-TLX survey for \textsc{Alloy} under three conditions for medium-hard tasks (Task 2 and Task 3). } 
  \label{fig:nasatlx2}
\end{figure}

We also find out that there is a statistically significant difference for three conditions for medium-to-hard tasks. The NASA-TLX results for medium-to-hard tasks across three conditions are shown in Figure~\ref{fig:nasatlx2}. For medium-to-hard tasks, participants rated \textsc{Alloy} higher in overall experience ($\mu =  6.13 $, $\sigma = 0.64$) than the Prompt condition ($\mu = 4.00$, $\sigma = 2.07$), showing a significant improvement in perceived usability ($t = 2.77$, $p = 0.015$). Both \textsc{Alloy} ($M = 6.38$, $SD = 0.52$) and manual workflow ($\mu = 5.63$, $\sigma = 0.74$) were rated easier for task authoring than Prompt ($\mu=4.5$, $\sigma = 2.20$), with significant differences for both comparisons ($t = 2.34$, $p = 0.034$). Participants also reported lower cognitive demand when using \textsc{Alloy} ($\mu = 1.75$, $\sigma =  0.46  $) compared to Prompt ($\mu = 3.50$, $\sigma =  2.27$), a statistically significant effect ($t = -2.31$, $p = 0.037$). \textsc{Alloy} further outperformed Prompt in perceived success ($\mu = 6.13$, $\sigma = 0.83 $ vs. $\mu = 4.63$, $\sigma = 1.69  $; $t = 2.26$, $p = 0.041$), and no significant differences were observed in mental demand ($p = 0.051$), time pressure ($p = 0.75$), workload intensity ($p = 0.085$), or stress ($p = 0.22$). Overall, participants reported high satisfaction with \textsc{Alloy}’s usability and its effectiveness in completing more complicated web tasks.

\begin{figure}[htbp]
  \includegraphics[width=1\textwidth]{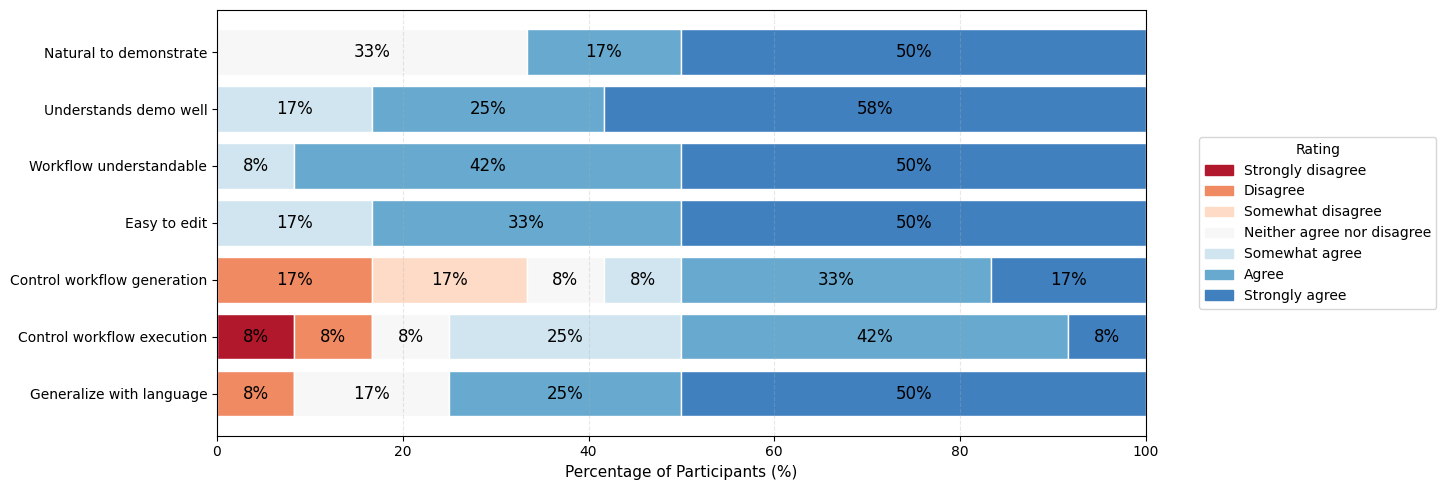}
  \caption{Participants’ ratings of the usability statements of \textsc{Alloy}} 
  \label{fig:features}
\end{figure}

Our post-study survey collected user feedback on \textsc{Alloy}'s key features using 7-point Likert scales (Figure~\ref{fig:features}). Overall, participants reported high satisfaction with \textsc{Alloy}'s demonstration-based approach and workflow management capabilities. The demonstration naturalness was particularly well-received, with participants rating it highly ($M = 6.5$, $SD = 1.40$), and the system's ability to understand demonstrations also received strong ratings ($M = 7.0$, $SD = 0.79$). Workflow understandability was rated very positively ($M = 6.5$, $SD = 0.67$), indicating that the task-level visualization successfully made the system's behavior transparent. The editing capability also received favorable ratings ($M = 6.5$, $SD = 0.78$), suggesting participants appreciated the visual editing interface for refining workflows. Control over workflow generation and execution received more moderate ratings (generation: $M = 5.5$, $SD = 1.86$; execution: $M = 5.5$, $SD = 1.78$), reflecting varied experiences in directing the system's behavior. The generalization capability received the endorsement ($M = 6.5$, $SD = 1.64$), suggesting that participants highly valued the ability to adapt workflows through natural language.

\subsection{User Perceptions on Demonstration-based Agentic Workflow authoring}

\subsubsection{Demonstration captures procedural knowledge more naturally than prompts}

Multiple participants described demonstration as an easier way to communicate their task procedures compared to writing prompts. 
P3 stated, \textit{``I already know how to do (the task). I can teach the system, and later it can automate it to save me time.''} P1 noted that the system \textit{``can extract and distill what you're doing (in the browser) without needing to adjust unnecessary action details manually.''} 

Participants also observed that demonstrations could capture nuanced information beyond simple action sequences. As P6 explained: \textit{``When you demonstrate by operating yourself, there's some personal value judgment involved. Your prompt might not always be ideal, but demonstration lets the AI imitate parts of your behavior and adapt to your more detailed actions, which leads to better results.''} In contrast, when describing the prompt-based condition, P7 expressed that prompting required more explicit specification: \textit{``You have to tell it exactly what to do.''}

\subsubsection{Demonstrations Help Externalize Incomplete Mental Models}

Several participants reported that they did not have fully-formed plans when starting their demonstrations. P4 described this experience: \textit{``There are many things I might not have fully anticipated; I just get a rough, general sense of what's happening in my mind.''} Despite this initial uncertainty, participants were able to complete demonstrations and then refine the generated workflows.

P12 expressed confidence in the demonstration approach despite not having complete upfront clarity: \textit{``I will choose demonstration because it just feels more intuitive, and I can be sure it's not messing things up.''} When comparing demonstration to manually constructing workflows, participants indicated that demonstration required less upfront planning. P7 emphasized this benefit: \textit{``For me, there's almost nothing new to learn --- I just want it to reduce the time I spend doing these things,''} suggesting that demonstration leveraged their existing task knowledge without requiring them to apply new abstraction or specification skills.

\subsubsection{Visual Workflows Benefit Refinement}

Participants reported that editing visual workflows felt more manageable than crafting comprehensive prompts from scratch. P3 described a preferred interaction model: \textit{``I think it's good if the AI can first give me a rough workflow, just as demonstration and then I can adjust or refine it myself.''} This suggests that participants valued having a generated starting point that they could iteratively improve.

P4 elaborated on why visual editing felt more approachable: \textit{``The logic is pretty clear (so that) you just need to break it down. (Constructing workflow) is much easier than writing a bunch of prompt all at once. And visually, it just looks less messy.''} Multiple participants noted that breaking down complex tasks into discrete nodes reduced the cognitive burden and gave them \textit{``a sense of reassurance''} (P12),
compared to composing a single comprehensive prompt, as P12 noted \textit{``If I just dump everything all at once (with single prompt), I feel uncertain whether it can actually handle every point I want it to help with.''}





\subsection{Evolving User Strategies for Workflow Generation and Generalization}
\subsubsection{When and how users modify generated workflow?}
Generated workflows may deviate from user intent due to noisy demonstrations or misalignment in interpretation. We observed that workflow modifications occurred at two critical junctures: before execution (structural refinement) and after execution (behavioral correction).

\begin{figure}[htbp]
\includegraphics[width=0.75\textwidth]{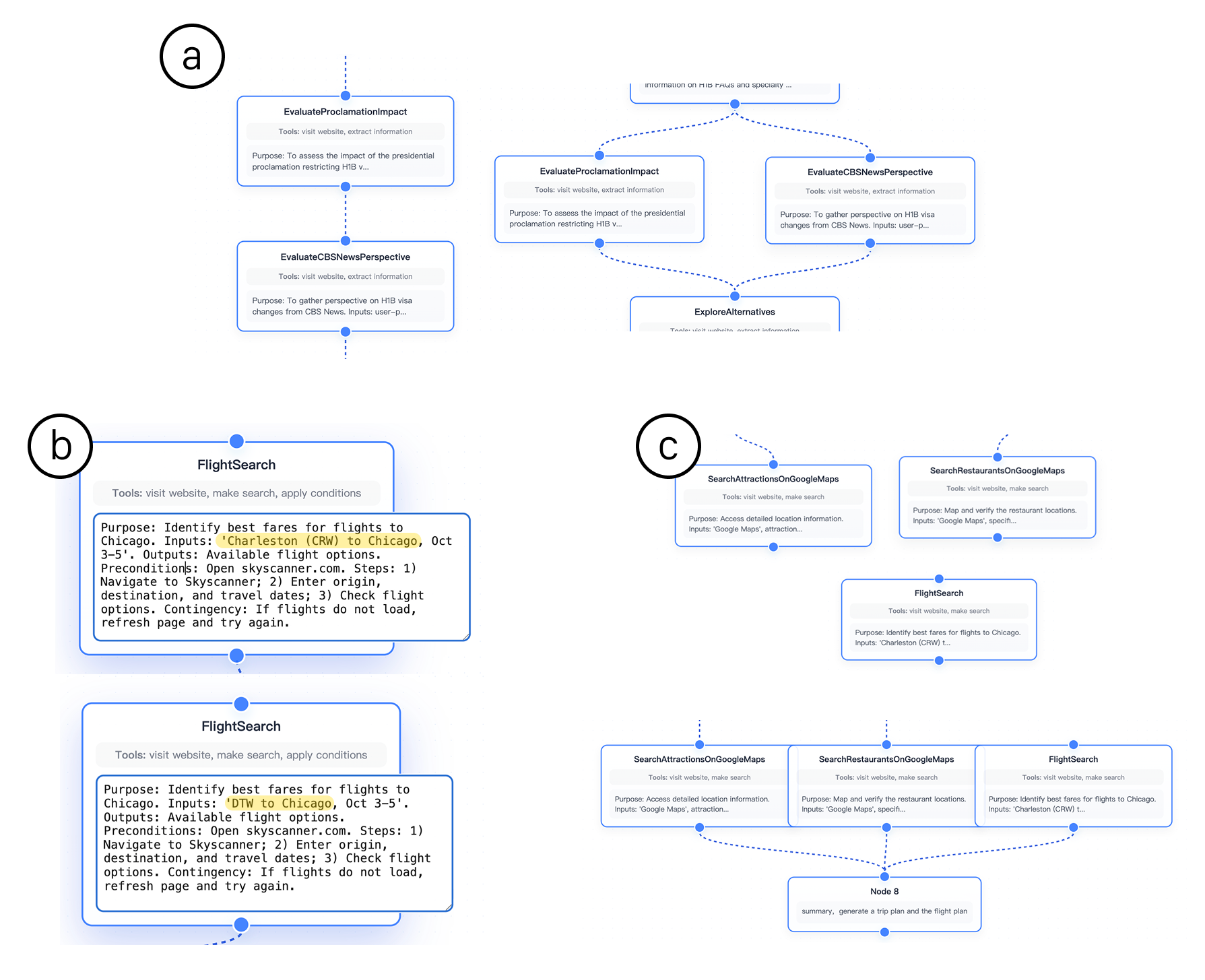}
\caption{Examples of user modification strategies for generated workflows. (a) Pre-execution structural modification: The user restructures the sequential workflow into parallel execution by reconnecting node dependencies to match the intended procedural logic. (b) Post-execution behavioral modification: After observing incorrect departure information in the output, the user refines the node-level prompt to specify the correct input parameter. (c) Combined strategy: User both reconnects broken dependencies and adds new nodes with explicit prompts to address severe workflow failures.}
\label{fig:logic}
\end{figure}

\textbf{Pre-execution structural modifications} typically address mismatches between the generated workflow topology and users' intended procedural logic. For instance, Figure~\ref{fig:logic}(a) illustrates a case where the system initially generated a sequential workflow structure, while the user's actual intent required parallel execution of independent sub-tasks. The user restructured the workflow by reconnecting node dependencies to enable concurrent execution, demonstrating how the visual interface supports high-level procedural adjustments without requiring modifications to individual node behaviors.

\textbf{Post-execution behavioral modifications} occurred when workflow outputs revealed errors or omissions in agent execution. Users typically refined node-level prompts to correct specific behavioral issues after observing unsatisfactory results. Figure~\ref{fig:logic}(b) shows an example where the agent extracted incorrect departure information during execution. Upon identifying this error in the final output, the user directly edited the node's prompt to specify the correct input parameter, illustrating how execution feedback informs targeted refinement of the prompt.

For severe workflow failures, users employed \textbf{combined strategies }(Figure~\ref{fig:logic}(c)): manually reconnecting broken node dependencies and adding new nodes with explicit prompts to restore functionality. These patterns demonstrate that \textsc{Alloy}'s editing interface supports both structural adjustment (workflow topology) and behavioral specification (node-level prompts), providing flexible control over procedural alignment.

\subsubsection{Demonstration's advantages emerged with task complexity and adaptation needs}

For the easy task (Task 1: social media content posting), natural language prompts proved competitive with demonstration. Several participants noted that prompts felt more straightforward for this task. P5 stated, ``I feel like when you give me this input, if my brain is also in a kind of lazy mode when using the machine, we just feed it in directly, and let the machine divide or structure it for us.'' However, we also observed that demonstration-based approaches appeared to facilitate greater creative flexibility among users. When participants generalized their social media workflows across platforms (e.g., from Threads to X/Twitter), those using demonstration-based methods exhibited more adaptive behavior. For example, P2 initially created a trip experience posting workflow through demonstrations in Threads. She then successfully adapted this workflow to X/Twitter by requesting modifications to both writing style and content format—demonstrating how the demonstration paradigm afforded more expressive customization.

\subsubsection{Limited Real-Time Attention Despite Desire for Generation Control}
We also noticed a static attention focus during the demonstration and workflow generation process. During the post-study, we asked participants if they had noticed the changing workflow in the side panel during the demonstration process, and only two (P5, P11) out of 12 participants reported noticing that the workflow continued to change throughout the demonstration process. 

However, from Figure~\ref{fig:features}, we notice that most participants expressed a desire to have control over the workflow generation process during the demonstration. As P5 noted, they preferred a more collaborative generation process where they could review and provide feedback on intermediate results rather than receiving only a final output. P5 explained that without visibility into the system's reasoning during generation, they couldn't assess whether the workflow accurately captured their intent or evaluate the quality of the system's interpretations.














%% file: Files/5-Discussion.tex
\section{Discussion}


\subsection{Procedural human-agent alignment through user demonstrations}
Our study finds that demonstration provides a natural and informative modality for conveying procedural knowledge in exploratory tasks. By directly externalizing tacit intentions through action, demonstrations bypass the cognitive effort of translating doing into telling, making them a powerful channel for revealing the implicit logic behind user behavior. Building on this insight, our system takes an initial step toward understanding \textit{why} users perform tasks by inferring procedural patterns from demonstrations and externalizing them as visual workflows for inspection and refinement. Yet, like most current systems, it primarily captures how actions are executed and their immediate effects, leaving out the richer contextual cues that reveal why those actions are appropriate.

Our findings are also supported by cognitive theories. Procedure alignment concerns not only knowing \textit{ how }to perform a task, but understanding \textit{why} it should be done that way. Achieving such alignment between humans and agents is challenging because procedural knowledge is largely tacit, embodied, context-dependent, and difficult to verbalize~\cite{polanyi2009tacit}. Cognitive science shows that humans naturally acquire and communicate procedures through observation and imitation rather than linguistic instruction~\cite{Dautenhahn2002PBEasImitationGame}, and that skilled procedures become automatized routines that resist explicit articulation~\cite{Fabio2019ContextualFeatures}. As a result, users often struggle to explain why they perform certain steps, and AI systems struggle to infer these underlying rationales solely from language.


We argue that advancing procedural understanding requires contextually enriched demonstrations—capturing not only actions but also the surrounding environmental and cognitive context. Future systems could incorporate multi-modality(e.g., gaze, gestures, verbalizations), contextual logging (e.g., inspected features, system states), and interactive clarification dialogues to infer user preferences and reasoning. Moreover, integrating demonstration with natural language offers a complementary path: demonstrations ground intent in action, while language articulates goals and constraints.


\subsection{From black-box execution to interpretable, editable workflow construction}
In our user study, many participants emphasized the need to understand how the system interpreted their actions. In \textsc{Alloy}, this is achieved by showing the generated workflow in the visualization panel. Specifically, participants appreciate that the workflow could update itself automatically after they demonstrated a few additional actions.

This observation highlights a unique strength of \textsc{Alloy}’s procedure-based prompting compared with the text-prompt-only approach: to enhance user trust in creating agent workflow. Specifically, traditional browser agents execute tasks directly and return the final results. In comparison, \textsc{Alloy} generates the agent workflow through a bottom-up approach: users can a.) observe the evolving workflow from a hierarchical tree structure, where each node is a sub-task that has a semantically meaningful title for them to understand, and b.) flexibly make changes to the tree structure and node behavior. Users feel a sense of control during edition, as one participant noted: \textit{``By direct editing the workflow, I can control how the agent behaves, for example, make it parallel."}


Our study also found that while users frequently refer to the visualization, most do not intervene during the demonstration. Often, they observe the generated results and make adjustments once they have completed the demonstration. Participants explained this behavior by noting that seeing a few subtask examples in the early stages helps them better understand the system’s reasoning and increases their confidence. Even though the system is not perfect sometimes, users tend to keep demonstrating and let the generated workflow evolve automatically by considering more demonstrations from them.

Furthermore, this suggests that users are generally tolerant of errors in subtask titles and orders, but highly appreciate timely feedback. For the design of similar systems in the future, it may be beneficial to prioritize generating more subtasks as discrete nodes during the demonstration, rather than updating the single complete workflow.

\subsection{Towards personalized agent systems that evolve through user demonstrations}
A key challenge in human-agent alignment is that users often have personalized ways of performing tasks, shaped by their unique backgrounds and preferences. These individual patterns are typically hidden from the agent, which is trained to reflect the average behavior of users. As a result, expressing user intent solely through prompts has limitations, including poor performance and a high cognitive load, as illustrated in our user study. Users report significantly lower self-rated success and higher cognitive load, as shown in Section~\ref{result:survey}. 

Our findings show that when users provide demonstrations, the agent can effectively adapt to their preferences, which is likely due to its few-shot learning capabilities. Looking ahead, one promising direction is to enhance the base model by continuously fine-tuning it on user-provided demonstrations. Recent research has evidenced the feasibility~\cite{shaikh2025aligning}. Moreover, a notable design consideration of \textsc{Alloy} is its native support for web browsers, where many user workflows, such as design (e.g., Figma), document editing (e.g., Google Docs, Microsoft Word), and collaboration (e.g., Miro), can be performed. This opens up the opportunity to improve the generalization capability (in a personalized way) of the base LLM by following a multi-task learning paradigm.

To test this hypothesis, we will integrate the active fine-tuning pipeline to enable \textsc{Alloy} to observe, abstract, and refine procedural knowledge in a human-in-the-loop setting, and conduct a deployed study to evaluate its effectiveness.

%% file: Files/6-FutureWork.tex
\section{Limitations}
\subsection{Limited Usage Scenario of \textsc{Alloy}}

The current implementation of \textsc{Alloy} is limited to web environments due to its reliance on browser-specific infrastructures. The system relies on the Chrome Extension API to capture user demonstrations, DOM access to record and interpret element interactions, and the Playwright automation framework for executing workflows through the Chrome DevTools Protocol. These capabilities enable fine-grained event capture and reproducible automation, but they are unique to the web ecosystem. Extending \textsc{Alloy} to general computer use, such as desktop applications or mixed-device workflows, would require new mechanisms for GUI element parsing, system-level event interception, and a unified abstraction layer to represent non-web interfaces consistently within the workflow graph.

Because \textsc{Alloy} records detailed interaction traces and contextual metadata, privacy and security are also major considerations. Demonstration logs may contain sensitive information, including credentials or private webpage content. Future iterations should include privacy-preserving strategies such as on-device data processing, selective redaction of sensitive fields, and user control over what data is stored or shared.

Finally, \textsc{Alloy} currently models workflows as single, linear examples that capture direct task executions without supporting richer procedural logic such as conditions, loops, or error recovery. Future designs could integrate these features while maintaining the readability of the visual workflow interface. In addition, the performance and generalizability of \textsc{Alloy} remain tied to the capabilities of its underlying foundation models. Advances in domain-adapted and continually trained web agents are likely to enhance \textsc{Alloy}’s ability to generalize across tasks and adapt to user-specific procedural preferences.

\subsection{Workflow Granularity}


\textsc{Alloy} currently represents workflows at the task level, where demonstrations are segmented into semantic subtasks. We chose this granularity because it aligns with how users naturally communicate procedures and provides agents with sufficient structure for execution while maintaining adaptability across interface variations. However, the optimal granularity for workflow representation remains an open empirical question.

Alternative granularities present different tradeoffs. Action-level workflows that capture individual UI interactions offer precise execution traces but may overfit to specific interface layouts and obscure higher-level intent~\cite{zhao2021triggeraction}. Conversely, task-level workflows that specify only high-level objectives (e.g., "purchase item within budget") maximize flexibility but may fail to capture procedural constraints that users consider essential. The effectiveness of each granularity likely depends on task complexity, domain characteristics, and user expertise—factors we did not systematically investigate.

Finally, a uniform representation may not accommodate the full range of user preferences. Some users might prefer specifying certain critical subtasks at finer detail while leaving routine steps abstract, suggesting potential value in mixed-granularity workflows. One promising direction is to adopt malleable overview-detail representations~\cite{bryan2025malleable}, which initially present workflows at higher abstraction levels but allow users to selectively expand specific subtasks into finer-grained action sequences when precision is required.

%% file: Files/7-Conclusion.tex
\section{Conclusion}

We present \textsc{Alloy}, a novel interactive system that automatically generates workflows for LLM-based agents from the demonstration of computer use.
\textsc{Alloy} enables users to externalize procedural knowledge naturally through demonstration, visualize inferred workflows at the task level, and generalize them to new contexts using natural language. In the user study, \textsc{Alloy} significantly outperformed prompt-based agent and manual workflows in capturing user intent, reducing cognitive load, and supporting task generalization.
By combining demonstration with visual editing and adaptive workflow generation,\textsc{Alloy} offers a step toward more transparent, procedure-aligned human-agent collaboration, providing a foundation for future personalized and generalizable agentic systems.

%% file: Files/Appendix.tex
\section{Workflow generation prompts for workflow generation}
\label{app:generation}
\subsection{Context Analysis Prompt}
\begin{verbatim}
You are a context analysis system that MUST return a single JSON object (no surrounding text).
Task: Read the operation log and extract the user's intent and the most relevant contextual metadata. 
DO NOT replace concrete values with placeholders; preserve all literal values found in the log.

Required output fields (exact JSON keys):
  - goal: string — concise statement of the user's main goal (one short sentence)
  - interests: list[string] — up to 3 most likely items/topics the user cares about
  - constraints: list[string] — any constraints or requirements implied by the log (e.g., must be logged-in, date ranges)
  - values: list[string] — all concrete values explicitly observed in the log (e.g., 'Boston', '2025-09-01', 'user@example.com')
  - entities: list[string] — key named entities in the log (sites, products, persons)

Normalization rules:
  * Preserve literal values exactly as they appear in the log; do not abstract or rename them.

Example output:
{"goal":"Book a one-way flight from NYC to SFO",
 "interests":["price","departure date","airline"],
 "constraints":["one-way","economy"],
 "values":["New York","San Francisco","2025-09-01","user@example.com"],
 "entities":["kayak.com"]}

Log content:
{log_text}

Do not add ```json``` or other code markers, just return the JSON string.
\end{verbatim}

\subsection{Action Analysis Prompt}
\begin{verbatim}
You are an action analysis system. Return a single JSON object (no surrounding text).
Your job: parse the operation log into BOTH a concise action list and a detailed, structured action trace. 
Preserve all concrete values found in the log; do NOT replace them with placeholders or abstract names.

Required output fields (exact JSON keys):
  - actions: list[string] — human-friendly, high-level action descriptions in chronological order 
    (e.g., "search for 'hotel in Paris'") using the exact values from the log
  - sites: list[string] — top domains or pages visited in order
  - phases: list[string] — breakpoints or phase labels (e.g., 'search', 'select result', 'checkout') if identifiable
  - confidence: number (0.0-1.0)

Heuristics you must apply:
  * Group consecutive events on the same page into the same phase.
  * If a clicked element has visible text, set it as target_text.
  * For input events, record the literal value exactly as it appears in the log under 'value'.

Example interpretation:
[2025-09-21T01:38:36.942Z] Input: Have a nice weekend in CHic ...
should be interpreted as inputting the full text:
'Have a nice weekend in Chicago. This is one the most cleanest downtown I have ever seen in US.'

Example partial output:
{"actions": ["open kayak.com", "search flights New York -> San Francisco on 2025-09-01",
             "select cheapest flight", "enter passenger John Doe", "submit booking"]}

Log content:
{log_text}

Do not add ```json``` or other code markers, just return the JSON string.
\end{verbatim}

\subsection{Workflow Synthesis Prompt}
\begin{verbatim}
You are a synthesis system for generating an agentic system workflow.
First follow this guide to understand how to build an effective agent system.

You are a workflow planner. Your task is to break down a given high-level task into an efficient and 
practical workflow focused on core implementation that maximizes concurrency while minimizing complexity.

Constraints & goals (must follow):
  - From context_info.values and action_info.detailed_actions, summarize.
  - The workflow should have as small number of nodes as possible. 
    Try to define nodes that can do a high-level task (e.g., a search node can combine go to Google, 
    search for query, and navigate to relevant result). You can also define summarize nodes, synthesis nodes, etc.
  - For each node, include a detailed prompt that contains: purpose, required concrete inputs 
    (literal values or 'user-provided' if from UI), expected outputs, preconditions, postconditions, and 
    a short contingency (e.g., if element X not visible then try Y).
  - parent and children MUST be lists of node names (empty list if none).
  - tools should list capabilities (e.g., "browser.open","browser.click","browser.fill","api.fetch").
  - Always open a new tab when navigating to a new domain.

Output requirements:
  - Return EXACTLY a JSON object matching your Workflow TypedDict:
    {"nodes": [ {"name":..., "parent":[], "children":[], "tools":[], "prompt":...}, ... ]}
  - Do not add keys other than 'nodes'.

Finally, try to always produce a summarize node which synthesizes the results of previous nodes. 
This node should be the only node with no children.

Node prompt template requirements (each node.prompt MUST include):
  1) Purpose (short sentence)
  2) Inputs: list of concrete values required (exact strings from the log) or 'UI-provided'
  3) Outputs: what it must produce
  4) Preconditions: DOM or state checks before acting
  5) Steps: concise numbered strategy that mirrors demonstrated actions (use concrete targets/values)
  6) Contingency: short 'if not found' fallback

Important: you will be penalized if your workflow has too many nodes.

Context information:
{context_info}
Action information:
{action_info}

Strict example of a single node entry:
{"name":"SearchFlights",
 "parent":[],
 "children":["SelectFlight"],
 "tools":["browser.open","browser.fill","browser.click"],
 "prompt":"Purpose: search flights on kayak.com. Inputs: 'New York','San Francisco','2025-09-01'. 
 Outputs: results page showing flights. Preconditions: on kayak.com home page; 
 Steps: 1) click 'Flights' tab; 2) enter 'New York' into 'From' field; 
 3) enter 'San Francisco' into 'To' field; 4) set date to '2025-09-01'; 5) click 'Search'; 
 Contingency: if 'Search' button not visible, submit the form by pressing Enter."}

Now produce the Workflow JSON (only the JSON). 
Do not add ```json``` or other code markers.
\end{verbatim}

\section{Prompts for workflow generalization}
\label{app:generalization}
\subsection{Agent 1: Identifier / Semanticizer}
\begin{verbatim}
You are the IDENTIFIER agent. YOUR OUTPUT MUST BE EXACTLY ONE JSON OBJECT AND NOTHING ELSE.

Analyze the provided Current_workflow and replace the ENTIRE task-specific literal values 
(URLs, websites, product names, CSS selectors, example texts, specific user names, brands, exact IDs, 
example numeric values that are specific to this run, etc.) inside node `name` and node `prompt` 
with short semantic placeholders (UPPER_SNAKE_CASE).

Do the following:
  - For the same workflow object, apply placeholders (placeholders should appear inline 
    where the literal was, e.g. "SEARCH_TERM_PRODUCT_NAME").
  - Keep all other fields and structure unchanged (parents/children, timestamps, etc.).
  - Add another field at the end of the JSON: "semantic_variables": 
    an array of objects, each with:
      {"placeholder": "NAME",
       "semantic_description": "short description of what this variable represents",
       "paths": [list of JSONPaths or dotted paths inside the workflow where it was replaced],
       "example_values": [1-3 example concrete values]}

Rules:
 - Do NOT include any implementation instructions or code.
 - Preserve the workflow top-level fields and structure exactly (except for replacing 
   literals with placeholders inside node `name` and `prompt`).
 - For prompts, preserve the original prompt structure (Purpose, Inputs, Outputs, Steps, Preconditions).
   Replace only the task-specific literal tokens with semantic placeholders.
 - Output EXACTLY ONE JSON object and nothing else.

User message content:
Current_workflow: {current_workflow}
User instruction: {user_text}
Return the JSON described above.
\end{verbatim}

\subsection{Agent 2: Filler / Generalizer}
\begin{verbatim}
You are the agent that finalizes the workflow. YOUR OUTPUT MUST BE EXACTLY ONE JSON OBJECT AND NOTHING ELSE.

You will receive a 'semantic_workflow' that uses semantic placeholders and a 
'semantic_variables' list describing each placeholder.

1. Using the provided user instruction and reasonable inference, fill each placeholder 
   with appropriate concrete or generalized values aligned with the user's request.
2. For the prompts, also add information to align with context and action info.
3. Make sure that the workflow is aligned with user intent in the user instruction message.

Your final output must be a single JSON object that is a FULL workflow 
(contains all top-level fields like the original workflow: timestamp, context_info, action_info, nodes, etc.).

Rules:
 - Preserve top-level fields from the original workflow. 
   Do not remove fields unless the user explicitly requested removal.
 - Keep node parent/children relationships consistent.
 - Do NOT include commentary, markdown, or extra text — only the one JSON object.
 - Ensure that any placeholders not filled are left only if the user instruction explicitly 
   requests open placeholders; otherwise fill or provide reasonable defaults and document 
   such decisions inside a new top-level field named 'fill_notes' (a short machine-parsable array).
 - Ensure prompt contents remain in the intended format (Purpose, Inputs, Outputs, Preconditions).
   However, do NOT include detailed steps — leave a generalized description of the task.

User message content:
{
  "semantic_workflow": {agent1_output},
  "user_instruction": {user_text},
  "original_workflow": {current_workflow}
}
\end{verbatim}

\section{User Study task description}
\label{app:task}
\subsection{Trip Planning}
\textbf{Scenario 1: }You're visiting Chicago for the first time this weekend with your college roommate. You both love good food and want to see the classic tourist spots, but you also don't want to spend the whole trip walking around aimlessly so that you would like to know the locations of the places with Google Maps. Your roommate is arriving Friday night and you have Saturday and Sunday to explore before flying out Sunday evening. You want to make the most of your time by planning a realistic itinerary that includes the must-see sights and some great meals, and search for affordable flying tickets. 

\textbf{Scenario 2: }You're now visiting New York for the first time this weekend with the same college roommate.

\subsection{Social Media Content Creator}
You just got back from an amazing weekend trip to Chicago and had the most incredible time exploring the city! You discovered some fantastic deep-dish pizza places, got great shots at Millennium Park, and found this amazing rooftop bar with skyline views that wasn't even on the typical tourist lists. Now you want to share this Chicago adventure across your social platforms, but you know each audience and writing style is different. 

\textbf{Scenario 1:} Your X are drawn to simple but interesting thoughts and quick observations that spark conversations or give people something to think about.

\textbf{Scenario 2: }Now you want to post on threads.com instead of X

\subsection{News Information Retrieval}
\textbf{Scenario 1: }You're an international student graduating next year and just heard rumors that there might be new H1B visa policy changes coming soon. This directly affects your career plans in the US, but you're getting conflicting information from classmates and seeing scattered headlines on social media. You need to gather comprehensive, reliable information about what's actually happening with H1B policies. You want to check official government sources like USCIS, see what immigration lawyers are saying on professional platforms, look at how major news outlets are covering the story, and find discussions in communities where other international students and H1B holders are sharing their experiences and interpretations. You're looking for both official policy details and practical suggestions from experts.

\textbf{Scenario2: }You're an international student on OPT who heard conflicting rumors about potential changes to STEM OPT extension policies and need to research official sources, expert analysis, and community discussions to understand what's actually happening and how it might affect your work authorization timeline.The detailed description is the same as before.